\documentclass[conference]{IEEEtran}
\usepackage{booktabs}
\usepackage{color}
\usepackage[table]{xcolor}
\usepackage{subcaption}
\usepackage{listings}
\usepackage{mdframed}
\usepackage{multirow}
\usepackage{tcolorbox}
\usepackage{xcolor} 
\usepackage{mdframed} 
\usepackage{hyperref}
\usepackage{float}

\definecolor{light-gray}{gray}{0.95}
\newcommand{\tool}{\texttt{jscefr}}

\begin{document}

\title{\texttt{jscefr}: A Framework to Evaluate \\ the Code Proficiency for JavaScript}

\author{\IEEEauthorblockN{Chaiyong Ragkhitwetsagul\textsuperscript{*}, Komsan Kongwongsupak\textsuperscript{*},
Thanakrit Maneesawas\textsuperscript{*}, Natpichsinee
Puttiwarodom\textsuperscript{*},\\ Ruksit
Rojpaisarnkit\textsuperscript{$\dagger$}, Morakot Choetkiertikul\textsuperscript{*}, Raula Gaikovina Kula\textsuperscript{$\ddagger$}, Thanwadee Sunetnanta\textsuperscript{*}}
\IEEEauthorblockA{\textsuperscript{*}Faculty of Information and Communication Technology (ICT), Mahidol University, Nakhon Pathom, Thailand\\
\textsuperscript{$\dagger$}Graduate School of Science and Technology, Nara Institute of Science and Technology, Nara, Japan\\
\textsuperscript{$\ddagger$}Graduate School of Information Science and Technology, Osaka University, Osaka, Japan
}
}
\maketitle

\begin{abstract}
In this paper, we present \tool~(pronounced \texttt{jes-cee-fer}), a tool that detects the use of different elements of the JavaScript (JS) language, effectively measuring the level of proficiency required to comprehend and deal with a fragment of JavaScript code in software maintenance tasks. Based on the pycefr tool, the tool incorporates JavaScript elements and the well-known Common European Framework of Reference for Languages (CEFR) and utilizes the official ECMAScript JavaScript documentation from the Mozilla Developer Network. \tool~categorizes JS code into six levels based on proficiency. \tool~can detect and classify 138 different JavaScript code constructs. To evaluate, we apply our tool to three JavaScript projects of the NPM ecosystem, with interesting results. 

A video demonstrating the tool's availability and usage is available at~\url{https://youtu.be/Ehh-Prq59Pc}.
\end{abstract}



\section{Introduction}
\label{intro}

JavaScript has consistently been one of the most popular programming languages used worldwide over the last five years. According to the 2023 GitHub Octoverse report~\cite{GitHub}, developers employed nearly 500 primary languages to build software on the platform. Despite the competition, JavaScript has consistently reigned supreme as the top language on GitHub from 2014 to 2022. Furthermore, a survey by Stack Overflow~\cite{rankingStackoverflow} indicates that JavaScript has been the most commonly used programming language for ten years in a row. One reason could be the massive NPM ecosystem of third-party libraries and the language's flexibility, which allows for fast development and usage in web applications.

Understanding coding proficiency is essential for developers as it drives career advancement, enhances problem-solving skills, increases efficiency and productivity, and ensures high-quality work. Proficient coders adapt more easily to new technologies, communicate effectively with their peers, and enjoy greater job security. Additionally, mastering coding skills fosters personal satisfaction, encourages innovation, and enables meaningful contributions to the developer community. 

Several methods have been proposed for evaluating programming language skills and proficiency. Sultana's study \cite{Sultana2016} outlined various criteria for assessing proficiency in programming as part of a Computer Science curriculum. This includes essential skills such as problem-solving, critical thinking, and the ability to apply logical reasoning within the context of Software Engineering. For instance, fundamental proficiencies involve explaining and utilizing basic data structures like arrays, linked lists, and dictionaries for practical tasks, classified as Level 1. In contrast, understanding advanced data structures such as B-trees, binomial and Fibonacci heaps, AVL/Red Black trees, splay trees, skip lists, and tries constitutes a Level 3 proficiency.
Additional research, such as the study by Ibezim et al. \cite{Ibezim2016ComputerPC}, focuses on the proficiencies necessary for computer education graduates to achieve sustainable employment in Enugu, Nigeria. This study highlights the importance of a balanced skill set, including hard skills, business acumen, and soft skills, outlining specific proficiencies needed in each category: 25 in hard skills, 18 in business skills, and 19 in soft skills. These are vital for graduates aiming to secure and maintain employment in software development.

Recently, there has been a focus on coding proficiency, particularly for the Python programming language. Several studies have examined the usage of different code constructs in Python and their effect on software development~\cite{Alexandru2018, Phan-udom2020, Sakulniwat2019, Leelaprute2022}. A study by Robles et al.~\cite{Robles2022} investigates the application of the Common European Framework of Reference for Languages (CEFR) to Python through code analysis. They developed a tool called pycefr, which analyzes a Python project and outputs the proficiency levels of the code constructs within the project.

CEFR~\cite{cefr} is an international standard for measuring language proficiency created to describe language learners' proficiency in speaking, reading, listening, and writing. It is one of the widely used Council of Europe policy instruments. It contains six-point scales ranging from beginner (A1), elementary (A2), intermediate (B1), upper-intermediate (B2), advanced (C1), and proficiency (C2). CEFR is a broad concept that can be applied to English and also other languages.
Different from Python, JavaScript has a well-documented history tracing back to ECMAScript\footnote{\url{https://tc39.es/ecma262/}}, which has grown to be one of the world's most widely used general-purpose programming languages. JavaScript is best known as the language embedded in web browsers but has also been widely adopted for server and embedded applications.

\begin{table}[]
{
\centering
\caption{Extracted from the MDN, these are the questions that reveal the skills needed for JavaScript Proficiency}
\label{tab:jscomplevels}
\resizebox{\columnwidth}{!}{ 
\begin{tabular}{p{1.5cm}p{7cm}}
\toprule
Level & Skills \\
\midrule
A1  &Can create and use variables\\
(Beginner) &Can perform mathematical calculations\\
&Can manipulate strings and arrays\\
&Can use conditional and loop statements\\
&Can create and use basic functions and events\\ 
\midrule
A2  & Can handle JSON data \\
(Elementary) &Can perform type conversion \\
&Can create and use literals \\
&Can use control flows and error handling\\
&Can use an advanced function, expression and operator, number and date, and string features \\
&Can manipulate collections \\
&Can use advanced class and object features\\ 
\midrule
B1 &Can use Iterations and Generators\\
(Intermediate) & Can use Meta Programming\\
&Can use JavaScript Modules\\ 
\midrule
B2  & Can use asynchronous programming\\
(Upper- & Can use client-side web APIs\\
Intermediate) & Can use Frameworks\\ 
\midrule
C1  & Can use advanced language constructs such as a sparse array, the nested function \\
(Advanced) &Can handle complicated data types and data structures such as coercion and closure  \\  
\midrule
C2  & Can use inheritance with prototype chain \\ 
(Proficiency) & Can use typed arrays \\ 
& Can perform memory management\\ 
& Can use strict mode \\ 
\bottomrule
\end{tabular}
}
}
\end{table}

\begin{table*}[t]
{
\centering
\caption{Sample \tool~code constructs that map between the MDN four categories classes and the CEFR levels}
\label{tab:examples}
\resizebox{\textwidth}{!}{ 
\begin{tabular}{p{6cm}lllc}
\toprule
Construct & MDN Group & MDN Subclass & MDN Class & Level \\
\midrule
const & - & Basics & Complete Beginner & A1 \\
\texttt{var}     & -   & Basics    & Complete Beginner & A1 \\
Variable Assignment (=)     & -   & Basics    & Complete Beginner & A1 \\
\midrule
anonymous function  & Function & Building Block  & Complete Beginner & A2  \\
JSON & JSON & JavaScript Object & Complete Beginner & A2  \\
\texttt{Try Catch} & Control Flow and Error Handling & - & JavaScript Guide & A2  \\
\midrule
declaring class & Using Class & - & JavaScript Guide & B1  \\
Dot Notation (\texttt{obj.prop}) & Object Basics & JavaScript Object & Complete Beginner & B1  \\
\texttt{this} & Object Basics & JavaScript Object & Complete Beginner & B1  \\
\midrule
Using \texttt{Promise.all()} & Using Promises & - & JavaScript Guide & B2  \\ 
async await & Asynchronous Programming & Language Overview & Intermediate & B2  \\ 
Create and append new element & Manipulating Documents & Client-side Web API & Intermediate & B2  \\ 
\midrule
Offline assets storage & Client-Side Storage & Client-side Web API & Intermediate & C1 \\ 
closure (return function in function) & Function & - & JavaScript Guide & C1  \\ 
Primitive Coercion (ex. \texttt{Date}) & Type Coercion (Auto convert type) & Data type and Data structure & Intermediate & C1  \\ 
\midrule
Using canvas 3d & Drawing graphic on web & Client-side Web API & Intermediate & C2  \\ 
Proxies & Meta Programming & - & JavaScript Guide & C2  \\ 
\texttt{WeakRefs} & Data structure aiding memory management & Memory Management & Advanced & C2  \\ 
\bottomrule
\end{tabular}
}}
\end{table*}

In this tool paper, we introduce a prototype \texttt{jscefr}, which is an extension of \texttt{pycefr}~\cite{Robles2022}. Using the same level measurement as CEFR, each JavaScript construct can be classified into one of the six levels (i.e., A1--C2). First, we leverage the existing documentation from the official ECMAScript programming language, provided by the Mozilla Developer Network. Hence, \tool~introduces a novel process for creating JavaScript proficiency levels based on the Mozilla Developer Network (MDN) documentation, which are then verified by JavaScript experts. The contribution of this paper is the \tool~tool, which will open more opportunities to study JavaScript projects, a completely different ecosystem from Python.
This work is a starting point for understanding how coding proficiency can be characterized and utilized for software maintenance tasks. 

Our tool prototype and source code is made publicly available at \url{https://github.com/MUICT-SERU/jscefr}.

\section{MDN Web Docs}
\label{sec:mdn}
Previously known as the Mozilla Developer Network (MDN) and formerly Mozilla Developer Center, it is a documentation repository and learning resource for web developers\footnote{\url{https://developer.mozilla.org/en-US/}}. It was started by Mozilla in 2005 as a unified place for documentation about open web standards, Mozilla's own projects, and developer guides. MDN Web Docs content is maintained by Mozilla, Google employees, and volunteers (a community of developers and technical writers). It also contains content contributed by Microsoft, Google, and Samsung, who, in 2017, announced they would shut down their own documentation projects and move all their documentation to MDN Web Docs\footnote{\url{https://github.com/mdn}}. Topics include HTML5, JavaScript, CSS, Web APIs, Django, Node.js, WebExtensions, MathML, and others.

According to Wikipedia\footnote{\url{https://en.wikipedia.org/wiki/MDN_Web_Docs}}, the advantage of JavaScript is its rich history of documentation. In 2019, Mozilla started beta testing a new reader site for MDN Web Docs, which was launched on December 14, 2020. Since then, all editable content has been stored in a Git repository hosted on GitHub, where contributors open pull requests and discuss changes. On January 25, 2021, the Open Web Docs (OWD) organization was launched as a non-profit fiscal entity to collect funds for MDN development. As of March 2023, the top financial contributors of OWD are Google, Microsoft, Igalia, Canva, and JetBrains.

\begin{figure*}[h]
\centering
\includegraphics[width=1.8\columnwidth]{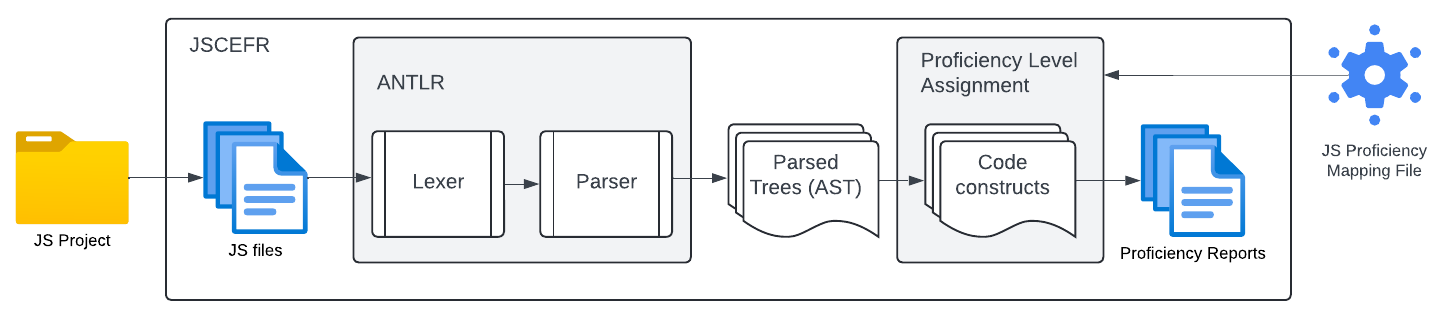}
\caption{\tool~System Overview}
\label{fig:sysarch1}
\end{figure*}

\section{Mapping the MDN and the CEFR}
\label{sec:creatingjscefr}
We followed the concept of CEFR by initially listing the skills that JavaScript developers can achieve at each proficiency level based on the examples and code constructs in MDN. We adopted the six proficiency levels (A1, A2, B1, B2, C1, C2) defined in CEFR but applied them to JavaScript programming. Our mapping team included the authors and a JavaScript expert with 8 years of experience in writing JavaScript, who reviewed our six proficiency levels and made adjustments accordingly.

The authors started by studying the groupings and topics in the Mozilla Developer Network (MDN\footnote{\url{https://developer.mozilla.org/en-US/docs/Web/JavaScript/About_JavaScript}}). In MDN, there were four categories: \textit{Complete Beginner}, \textit{JavaScript Guide}, \textit{Intermediate}, and \textit{Advanced}.

\begin{itemize}
    \item \textit{Complete Beginner} -- includes basic web programming, including basic JavaScript and client-side scripting.
    \item \textit{JavaScript Guide} -- provides fundamentals of JavaScript programming such as grammar, control flow, loops, and functions.
    \item \textit{Intermediate} -- explains more complex concepts such as client-side JavaScript frameworks, client-side web APIs, and data structures.
    \item \textit{Advanced }-- includes advanced concepts such as inheritance and the prototype chain, memory management, concurrency model, and event loop.
\end{itemize}
To assist with the classifications, we conducted a round-table discussion with all authors and defined questions that would be used to help aid how each construct would be assessed for each level. 
This is similar to the methodology for the CEFR levels definition\footnote{\url{https://www.coe.int/en/web/common-european-framework-reference-languages/table-1-cefr-3.3-common-reference-levels-global-scale}}, and is based on what a developer is able and not able to do. For example, we decided that a developer at level A2 should be able to create and manipulate a JSON data structure. 

Table~\ref{tab:jscomplevels} shows the JavaScript proficiency levels mapped to the CEFR levels. Levels A1 and A2 represent the basic skills that JavaScript developers possess, such as creating and using variables and handling JSON data. Levels B1 and B2 encompass more complex skills, such as using iterations and asynchronous programming. Lastly, levels C1 and C2 contain advanced skills, such as using nested functions and memory management.

\subsection{Mapping the JS constructs to CEFR levels}

The mapping of JS code constructs to CEFR levels is not trivial (i.e., we are mapping between six CEFR levels and four MDN classifications). Thus, careful discussion and analysis were required. We collected the code constructs shown on MDN in each of the categories and extracted them. Through round-table discussions, we decided and assigned an appropriate CEFR level, based on Table~\ref{tab:examples}, to the code constructs. In total, we gathered and assigned JavaScript proficiency levels to 180 code constructs.

\subsection{Feedback and refinement from Experts}

To ensure our JavaScript CEFR level assignments were valid, we interviewed 12 experienced JavaScript practitioners to verify our assignments. The criteria for experts who evaluated our JavaScript proficiency level assignments included having at least 4 years of experience in JavaScript and holding at least a senior developer position.

We randomly divided our 180 JavaScript constructs into 10 equal groups based on stratified sampling. The code constructs in each group were evaluated by two experts. If there were conflicts, we recruited another expert to make the final decision. The additional experts were recruited after the initial evaluation by the two experts had finished.

Table~\ref{tab:examples} shows the MDN groups, subclasses, and classes of the code constructs and the assigned CEFR levels.  The examples of the code constructs and their assigned proficiency levels (A1 to C2) are listed. From the table, we can see that basic code constructs such as const and try-catch are assigned to levels A1 and A2, respectively. More difficult code constructs, such as declaring classes and using async-await, are assigned to levels B1 and B2, respectively. Lastly, the most advanced code constructs, such as offline asset storage and proxies, are assigned to levels C1 and C2, respectively. Please note that, due to our validation with the experts, some of the code constructs were moved to different levels and do not follow the original groupings in MDN. After completing this step, we started implementing the \tool~tool.

\subsection{\tool~System Overview}
Figure~\ref{fig:sysarch1} shows the system overview of the \tool~tool.\tool~accepts a JavaScript project directory as input and outputs proficiency reports. The user can configure the proficiency levels by uploading their own \textit{JS proficiency mapping file} into the system, allowing for customized proficiency level assignments for each code construct. Once this process is completed, the analysis can be performed multiple times without needing to reconfigure.
\tool~then traverses the abstract syntax tree (AST) of the JavaScript files and extracts code constructs. Then, the proficiency level assignment is performed based on the given JavaScript proficiency mapping file.

\subsection{Increasing Detection Accuracy}
In our preliminary testing of the tool, we found that the code constructs retrieved from parsing the AST could only cover below 50\% of all code constructs in the proficiency configuration file. This is because ANTLR breaks the source code file into low-level code constructs, such as variable statements, if statements, and iteration statements, while some of our defined code constructs are at a more abstract level.

For example, \texttt{then()}, a function in JavaScript, belongs to the B2 proficiency level according to the proficiency matrix we arranged, but it is considered an identifier by the conventional ANTLR AST walker, which has no assigned level. To address this, we expanded ANTLR's parser capabilities by including more rules to be detected by the AST rule visitor and overwriting the visitor to detect these missing code constructs.

After this modification, the code construct coverage rose from 50\% to 75\% of all code constructs that we collected. The final \tool~tool contains 138 code constructs with complete proficiency assignments.

\subsection{Our Prototype Walk-through Scenario}
\paragraph{Input}
To start analyzing a JavaScript project, the user inputs the path to the JavaScript project they want to analyze. The files in the project are then retrieved. \tool~selects only JavaScript files and sends each of them to be further analyzed by the lexer and parser. We employ ANTLR for the lexer and the parser (Figure~\ref{fig:sysarch1}).

\paragraph{Output}
After executing the \tool~tool, the analysis results are shown in the command line (Figure\ref{fig:command_line}). Additionally, \tool~saves all the detected code constructs with the information of the directory name, file name, detected proficiency level, starting line, ending line, starting column, and ending column into a CSV file (as shown inFigure~\ref{fig:CSV_report}). 
The tool also generates reports in JSON format. 

\begin{figure}[tb]
\centering
\begin{lstlisting}
    ============================
    RESULT OF THE ANALYSIS:
    Analyzed .js files: 2
    Elements of level A1: 61
    Elements of level A2: 6
    Elements of level B1: 20
    Elements of level B2: 5
    ============================
\end{lstlisting}
\caption{Result after executing \tool}
\label{fig:command_line}
\end{figure}

\begin{figure}[tb]
\centering
\begin{lstlisting}[backgroundcolor=\color{light-gray},basicstyle=\scriptsize\ttfamily]
Repo,File,Class,Level,StartLine,StartCol,EndLine, EndCol
App,App/app.js,comment,A1,7,2,7,30
App,App/app.js,arrayLiteral,B2,1,15,1,62
App,App/app.js,elelmentList,A2,1,16,1,53
App,App/app.js,memberDotExpression,B1,2,12,2,21
App,App/app.js,querySelector,A1,3,23,3,23
App,App/app.js,memberDotExpression,B1,5,0,5,4
\end{lstlisting}
\caption{The CSV report}
\label{fig:CSV_report}
\end{figure}

\section{Initial Analysis on Real-World \\ Open Source Projects}
\label{sec:result}
We performed a preliminary evaluation of \tool~by analyzing three popular front-end JavaScript frameworks: Next.js, React, and SvelteKit. As shown in Table~\ref{tab:stats}, we analyzed the following versions:

\begin{itemize}
    \item Next.js version 13.1.1 (released on 24 December 2022)
    \item React version 18.2.0 (released on 22 June 2022)
    \item SvelteKit version 1.12.0 (released on 17 March 2023)
\end{itemize}
where Next.js had the largest number of JavaScript files (4,812), followed by React (1,859) and SvelteKit (596).

The experiment was conducted on a scalable Ubuntu 22.04 cloud machine with Intel Xeon processors (2.3 GHz), 16 GB of RAM, and 200 GB of disk space. \tool~successfully analyzed all three projects.

\begin{quote}
    \textit{``\tool~is able to run on real-world projects that may scale from 596 to 4,812 files.''}
\end{quote}

\begin{table}
\centering
\caption{Statistics of the analyzed projects}
\label{tab:stats}
\small
\begin{tabular}{llr}
\toprule
Project & Version & Number of JS files \\
\midrule
Next.js & 13.1.1 & 4,812 \\
React &  18.2.0 &  1,859 \\
SvelteKit & 1.12.0 & 596 \\
\bottomrule
\end{tabular}
\end{table}

\begin{table}
\centering
\caption{File-based proficiency of the analyzed projects}
\label{tab:files}
\small
\begin{tabular}{crrr}
\toprule
Proficiency Level & Next.js & React & SvelteKit \\
\midrule
A1 & 7 & 5 & 7 \\
A2 & 11 & 12 & 1 \\
B1 & 3,554 & 1,207 & 359 \\
B2 & 1,146 & 453 & 190 \\
C1 & 94 & 173 & 36 \\
C2 & 0 & 9 & 3 \\
\midrule
Total & 4,812 & 1,859 & 596 \\
\bottomrule
\end{tabular}
\end{table}

\begin{figure*}
     \centering
     \begin{subfigure}[b]{0.29\textwidth}
         \centering
         \includegraphics[width=\textwidth]{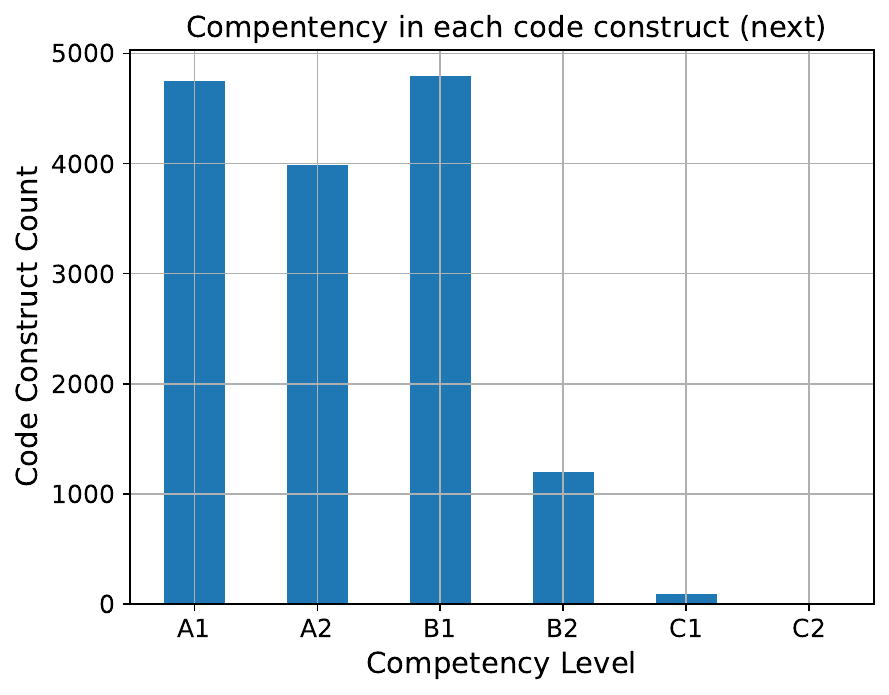}
         \caption{Next.js}
         \label{fig:next_hist_all}
     \end{subfigure}
     \begin{subfigure}[b]{0.29\textwidth}
         \centering
         \includegraphics[width=\textwidth]{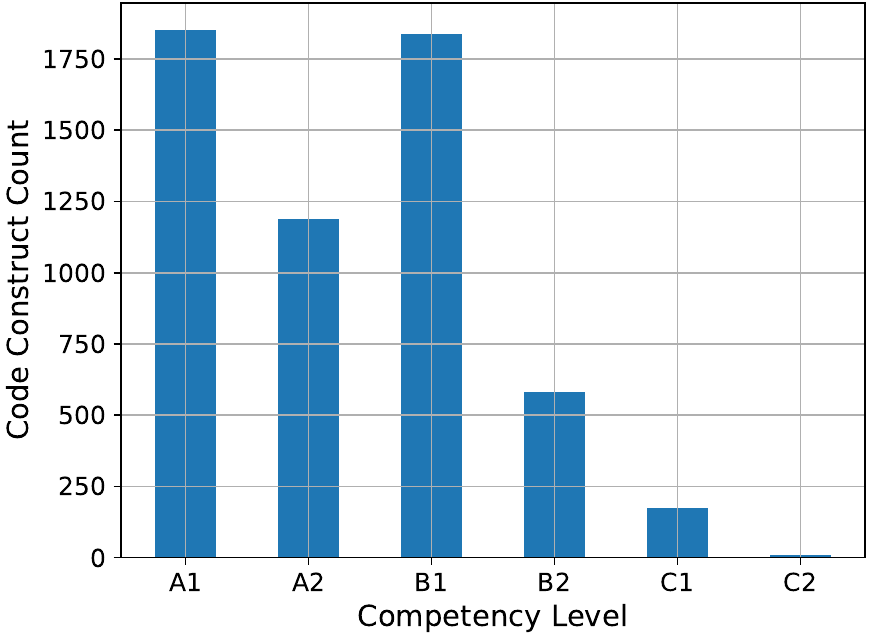}
         \caption{React}
         \label{fig:react_hist_all}
    \end{subfigure}
    \label{fig:react_hist}
    \begin{subfigure}[b]{0.29\textwidth}
         \centering
         \includegraphics[width=\textwidth]{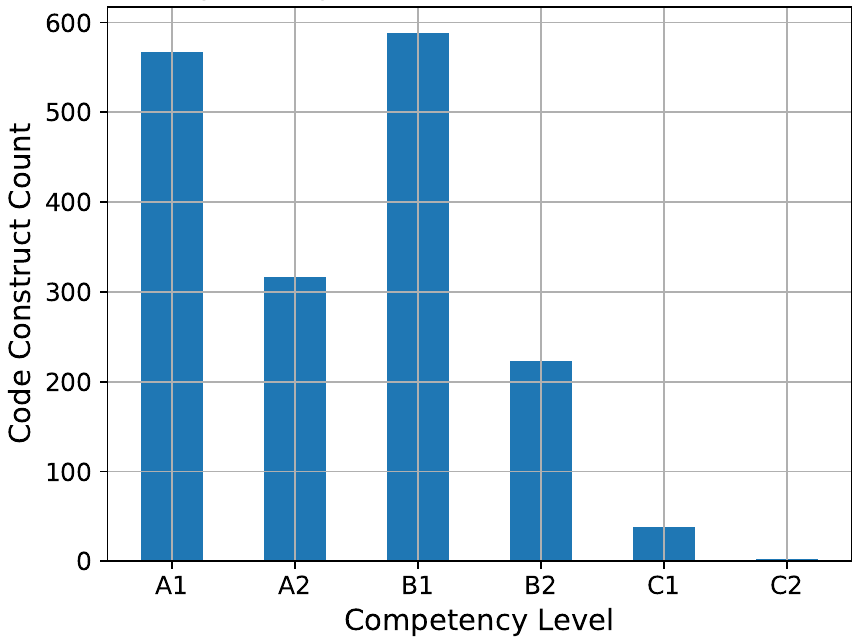}
         \caption{SvelteKit}
         \label{fig:SvelteKit_hist_all}
    \end{subfigure}
    \caption{Distributions of JavaScript proficiency levels in the analyzed projects}
    \label{fig:hist}
\end{figure*}

The analysis results of JavaScript code proficiency in the three projects by \tool~are shown in Table~\ref{tab:files} and Figure~\ref{fig:hist}. From Table~\ref{tab:files}, we present the classifications of files in the projects according to the highest proficiency level found in each file. 
This is because the developer who reads the file must understand the code construct with the highest proficiency level in order to understand all the code in the file.

We found that, for all three projects, the majority of the files have a proficiency level of B1 (3,554 for Next.js, 1,207 for React, and 359 for SvelteKit), followed by B2 and C1. Thus, the developers in these three projects must understand at least an intermediate level of JavaScript code constructs to maintain the projects. Only a few files have a beginner or advanced proficiency level of A1, A2, or C2.
\begin{quote}
    \textit{``We find that out of our three studied projects, only a total of 12 C2-level code proficient files were detected.''}
\end{quote}

Figure~\ref{fig:hist} presents a more fine-grained result of the number of JavaScript code constructs in the three projects divided by their proficiency levels. We found that the three projects have almost similar numbers of A1 and B1 code constructs, followed by A2 and B2. Regarding the advanced code constructs, Next.js has 94 C1 constructs and no C2 constructs. React has 174 C1 and 9 C2 constructs. Lastly, SvelteKit has 38 C1 and 3 C2 constructs.
This analysis shows that \tool~is scalable and can be applied to analyze real-world JavaScript projects.

\section{\tool~for Software Maintainance}
\label{sec:applications}
We foresee several potential studies and applications for \tool~and list two immediate applications as follows.
First, \textit{Assessing proficiency to complete a code review:}~One can adopt \tool~to study the proficiency of JavaScript code changes that are being reviewed. \tool~can extract and report the level of code constructs in the code changes, and recommendations for appropriate reviewers can be made based on the reported code proficiency.
Second, \textit{Assess open issues to assign to newcomers to an OSS project:~}The results from \tool~can also be used to select the right tasks for new contributors in open-source projects. In scenarios where updates need to be introduced to the existing codebase, project maintainers can select tasks that involve JavaScript files with lower proficiency levels to make it easier for new contributors.

\section{Discussion and Future Plans}
\label{sec:conclusion}
We present an automated tool for JavaScript code proficiency analysis called \tool. The tool leverages the concept of the Common European Framework of Reference for Languages (CEFR) and utilizes the official ECMAScript JavaScript documentation from the Mozilla Developer Network to define the proficiency levels of JavaScript code constructs.
Future work also includes several validations of our mappings, especially since we view proficiency levels as a sliding scale.
However, we are confident that our mappings are a starting point.
Furthermore, we plan to use \tool~for several empirical studies of JavaScript open-source projects as discussed in Section~\ref{sec:applications}, which also include comparing competencies across other different ecosystems like Python.  

\section{Acknowledgement}
This work is partially supported by Faculty of Information and Communication Technology, Mahidol University, and is supported by the Japanese Society for the Promotion of Science (JSPS) KAKENHI Grant Number JP20H05706. We thank Varayut Lerdkanlayanawat for providing useful suggestions about JavaScript throughout the whole study.

\bibliographystyle{IEEEtran}
\bibliography{references}

\begin{thebibliography}{10}
\providecommand{\url}[1]{#1}
\csname url@samestyle\endcsname
\providecommand{\newblock}{\relax}
\providecommand{\bibinfo}[2]{#2}
\providecommand{\BIBentrySTDinterwordspacing}{\spaceskip=0pt\relax}
\providecommand{\BIBentryALTinterwordstretchfactor}{4}
\providecommand{\BIBentryALTinterwordspacing}{\spaceskip=\fontdimen2\font plus
\BIBentryALTinterwordstretchfactor\fontdimen3\font minus \fontdimen4\font\relax}
\providecommand{\BIBforeignlanguage}[2]{{%
\expandafter\ifx\csname l@#1\endcsname\relax
\typeout{** WARNING: IEEEtran.bst: No hyphenation pattern has been}%
\typeout{** loaded for the language `#1'. Using the pattern for}%
\typeout{** the default language instead.}%
\else
\language=\csname l@#1\endcsname
\fi
#2}}
\providecommand{\BIBdecl}{\relax}
\BIBdecl

\bibitem{GitHub}
GitHub, ``The top programming languages,'' \url{https://octoverse.github.com/2022/top-programming-languages}, 2022.

\bibitem{rankingStackoverflow}
\BIBentryALTinterwordspacing
StackOverflow. (2022) Developer survey. [Online]. Available: \url{https://survey.stackoverflow.co/2022/\#most-popular-technologies-language}
\BIBentrySTDinterwordspacing

\bibitem{Sultana2016}
S.~Sultana, ``Defining the competencies, programming languages, and assessments for an introductory computer science course,'' Ph.D. dissertation, Old Dominion University Libraries, 2016.

\bibitem{Ibezim2016ComputerPC}
N.~E. Ibezim and C.~I. Chibuogwu, ``Computer programming competencies required by computer education graduates for sustainable employment,'' \emph{Review of European Studies}, vol.~9, p. 106, 2016.

\bibitem{Alexandru2018}
C.~V. Alexandru, J.~J. Merchante, S.~Panichella, S.~Proksch, H.~C. Gall, and G.~Robles, ``{On the usage of pythonic idioms},'' in \emph{Proceedings of the 2018 ACM SIGPLAN International Symposium on New Ideas, New Paradigms, and Reflections on Programming and Software (Onward! '18)}, 2018, pp. 1--11.

\bibitem{Phan-udom2020}
P.~Phan-udom, N.~Wattanakul, T.~Sakulniwat, C.~Ragkhitwetsagul, T.~Sunetnanta, M.~Choetkiertikul, and R.~G. Kula, ``{Teddy: Automatic Recommendation of Pythonic Idiom Usage For Pull-Based Software Projects},'' in \emph{Proceedings of the IEEE International Conference on Software Maintenance and Evolution (ICSME '20)}, 2020, pp. 806--809.

\bibitem{Sakulniwat2019}
T.~Sakulniwat, R.~G. Kula, C.~Ragkhitwetsagul, M.~Choetkiertikul, T.~Sunetnanta, D.~Wang, T.~Ishio, and K.~Matsumoto, ``{Visualizing the Usage of Pythonic Idioms Over Time: A Case Study of the with open Idiom},'' in \emph{Proceedings of 10th International Workshop on Empirical Software Engineering in Practice (IWESEP '19)}, 2019, pp. 43--435.

\bibitem{Leelaprute2022}
P.~Leelaprute, B.~Chinthanet, S.~Wattanakriengkrai, R.~G. Kula, P.~Jaisri, and T.~Ishio, ``{Does Coding in Pythonic Zen Peak Performance? Preliminary Experiments of Nine Pythonic Idioms at Scale},'' in \emph{Proceedings of IEEE/ACM 30th International Conference on Program Comprehension (ICPC '22)}, 2022, pp. 575--579.

\bibitem{Robles2022}
G.~Robles, R.~G. Kula, C.~Ragkhitwetsagul, T.~Sakulniwat, K.~Matsumoto, and J.~M. Gonzalez-Barahona, ``pycefr: Python competency level through code analysis,'' in \emph{2022 IEEE/ACM 30th International Conference on Program Comprehension (ICPC '22)}, 2022, pp. 173--177.

\bibitem{cefr}
\BIBentryALTinterwordspacing
{Council of Europe}. (2023) {CEFR Companion Volume: Enhancing engagement in language education}. [Online]. Available: \url{https://www.coe.int/en/web/common-european-framework-reference-languages}
\BIBentrySTDinterwordspacing

\end{thebibliography}

\end{document}